\documentclass[prd,letterpaper,tightenlines,
showpacs,preprintnumbers,nofootinbib,floatfix]{revtex4}

\usepackage{amsmath,amssymb,amsfonts,graphicx,pifont,dcolumn}
\usepackage{epstopdf}
\usepackage{color}
\usepackage{cancel}
\usepackage{ulem}
\RequirePackage{slashed}

\newcommand{\bit}{\begin{itemize}}
\newcommand{\eit}{\end{itemize}}

\def\be{\begin{equation}}
\def\ee{\end{equation}}
\def\bea{\begin{eqnarray}}
\def\eea{\end{eqnarray}}

\begin{document}

\title{Scattering of charged particles off monopole-anti-monopole pairs.}

\author{ Vicente Vento }

\affiliation{Departamento de F\'{\i}sica Te\'orica - IFIC, Universidad de Valencia - CSIC, E-46100 Burjassot (Valencia), Spain.}

\author{Marco~Traini} 
\affiliation{INFN - TIFPA, Dipartimento di Fisica, Universit\`a degli Studi di Trento, Via Sommarive 14, I-38123 Povo (Trento), Italy}

\date{\today}

\begin{abstract}
The Large Hadron Collider is reaching energies never achieved before allowing the search for exotic particles in the TeV mass range. In a continuing effort to find monopoles we discuss the effect of the magnetic dipole field created by a pair of monopole-anti-monopole or monopolium on the successive bunches of charged particles in the beam at LHC.

\end{abstract}

\pacs{14.80.Hv, 12.90.+b, 12.20.Fv}

\maketitle

\section{Introduction}
The theoretical justification for the existence of classical magnetic poles, hereafter called
monopoles, is that they add symmetry to Maxwell's equations and explain charge quantisation. Dirac showed that the mere existence of a monopole in the universe could offer an explanation of the discrete nature of the electric charge.
His analysis leads to the Dirac Quantisation Condition (DQC) \cite{Dirac:1931kp,Dirac:1948um}

\begin{equation}
e g = N/2, \; N = 1,2,...,
\label{dirac}
\end{equation}
where $e$ is the electron charge, $g$ the monopole magnetic charge and we use natural units
$\hbar= c=1= 4 \pi \varepsilon_0=\frac{\mu_0}{4 \pi}$. In Dirac's formulation, monopoles are assumed to exist as point-like particles and quantum mechanical consistency conditions lead to  establish the magnitude of their magnetic charge. Monopole  physics took a dramatic turn when 't Hooft \cite{'tHooft:1974qc} and Polyakov \cite{Polyakov:1974ek} independently discovered that the SO(3) Georgi-Glashow model \cite{Georgi:1972cj} inevitably contains monopole solutions \cite{Shnir:2005xx}. These topological monopoles are impossible to create in particle collisions either because of their huge GUT mass \cite{'tHooft:1974qc, Polyakov:1974ek} or for their complicated multi-particle structure \cite{Drukier:1981fq}. In the case of low mass topological solitons they might be created in heavy ion collisions via the Schwinger process \cite{Gould:2017zwi}. For the purposes of this investigation we will adhere to the Dirac picture of monopoles, i.e., they are elementary point-like particles with magnetic charge $g$  determined by the Dirac condition Eq.(\ref{dirac}) and with unknown mass $m$ and spin.
These monopoles have been a subject of experimental interest since Dirac first proposed them in 1931. Searches for direct monopole production have been performed in most accelerators.
The lack of monopole detection has been transformed into a monopole mass lower bounds \cite{Abulencia:2005hb,Fairbairn:2006gg,Abbiendi:2007ab,Aad:2012qi}. 
The present limit is $m> 400$ GeV  \cite{Aad:2015kta,Lenz:2016zpj,Acharya:2014nyr,MoEDAL:2016jlb,Acharya:2016ukt,Acharya:2017cio} but experiments at LHC can probe much higher masses. 
Monopoles may bind to matter and we have studied ways to detect them by means of inverse Rutherford scattering  with ions  \cite{Kazama:1976fm,Vento:2018sog}.

Since the magnetic charge is conserved monopoles at LHC will be produced predominantly in  monopole-anti-monopole pairs (or monopolium)~\cite{Dougall:2007tt,Vento:2007vy,Baines:2018ltl,Epele:2012jn}. This magnetic charge-less pair, given the collision geometry, will produce a magnetic dipole field.  We discuss hereafter the scattering of charged particles on a magnetic dipole and will analyze later on how our results affect the particles of the successive bunches at LHC. This development therefore assumes that the monopoles are more massive than the beam particles and therefore their scattering does not affect the dynamics of the formation process.

\section{Scattering of charged particles by a magnetic dipole.}

Suppose that at LHC a monopole-anti-monopole pair is produced by any of the studied mechanisms \cite{Gould:2017zwi,Milton:2006cp,Dougall:2007tt,Epele:2012jn,Baines:2018ltl}.   If the pair is produced close to threshold the pair will move slowly away from each other in the interaction region.  These geometry will produce a magnetic dipole in the beam line which affects the particles coming in the successive bunches. We model  this scenario as the study of the scattering of a beam of  charged particles by a fixed magnetic dipole created by two magnetic charges separated by a fixed distance. We will discuss the peculiarities of monopolium,  as a bound state, in Section \ref{monopolium}.

\begin{figure}[htb]
\begin{center}
\includegraphics[scale= 0.4]{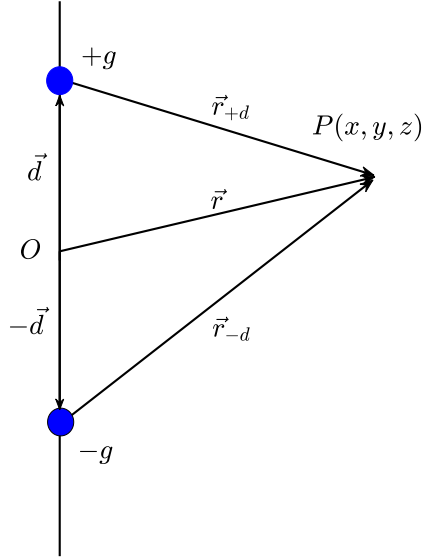} 
\end{center}
\caption{The magnetic moment configuration discussed in the text. } 
\label{magneticdipole}
\end{figure}

The magnetic field of a monopole located at the origin of the coordinate system is given by

\begin{equation}
\vec{B}_g = g \frac{\vec{r}}{r^3},
\label{Bmonopole}
\end{equation}
where $g$ is the magnetic charge,  $\vec{r}$ the radial vector of coordinates $(x,y,z)$ and $r$ the norm of $\vec{r}$. Let us construct the magnetic field of a monopole, located at position $\vec{d}= (0,0,d)$, and an anti-monopole (located at position $(0,0,-d)$), 
where $d$ is a distance (see Fig. \ref{magneticdipole}),

\begin{equation}
\vec{B}_d = g \frac{\vec{r}_{+d}}{r_{+d}^3} - g \frac{\vec{r}_{-d}}{r_{-d}^3},
\label{Bdipole}
\end{equation}
where $\vec{r}_{\pm d} = \vec{r} \mp \vec{d}$ and $r_{\pm d}$ their norm.

Let us perform an expansion in $d/r$. 

\begin{equation}
\vec{B}_d = 3 \frac{(\overrightarrow{\cal M} \cdot \vec{r})\; \vec{r}}{r^5} - \frac{\overrightarrow{\cal M}}{r^3} - \frac{3 d^2}{2 r^2}\left(5 \frac{(\overrightarrow{\cal M} \cdot \vec{r})\; \vec{r}}{r^5} - \frac{\overrightarrow{\cal M}}{r^3}\right) + \frac{5 d^2 z^2}{r^4}\left(7  \frac{ (\overrightarrow{\cal M} \cdot \vec{r})\; \vec{r}}{r^5} - 3 \frac{\overrightarrow{\cal M}}{r^3} + \ldots \right),
\end{equation}
where $\overrightarrow{\cal M}= 2g \vec{d}$ is the magnetic moment.
Note that the magnetic charge field vanishes in the expansion in $d$  as expected from duality and that to leading order in $d/r$ we obtain the conventional field for a fixed magnetic moment.

Let us study the behaviour of the vector potential which is important to determine the interaction between the charged particles of the beam and the magnetic moment. 
Duality hints us that once the singularities are taken care of the result should resemble the conventional case. The vector potential for a monopole whose magnetic field is Eq.(\ref{Bmonopole}) can be written as

\begin{equation}
\vec{A}_g= g \frac{1-\cos{\theta}}{r \sin{\theta}} \hat{\phi} = g \frac{r-z}{r (x^2+ y^2)}  (-y, x,0),
\label{Amonopole}
\end{equation}
where $\theta \in [0,\pi]$ is the spherical polar angle and $\hat{\phi} = (- \sin{\phi}, \cos{\phi}, 0)$  is the azimuthal unit vector, being  $\phi$ the azimuthal angle. Note that this field is singular for $\theta = \pi$, i.e. , this is the famous Dirac string singularity.
The  vector field generated by the monopole-anti-monopole of Fig. \ref{magneticdipole} is given by

\begin{equation}
\vec{A}_d = g \left(\frac{z+d}{\sqrt{x^2+y^2+ (z+d)^2}}  - \frac{z-d}{\sqrt{x^2+y^2+(z-d)^2}} \right) \frac{(-y, x,0)}{x^2+y^2}.
\label{vectord}
\end{equation}
If we perform a series expansion in $d/r$,  we obtain
\begin{equation}
\vec{A} = \frac{\overrightarrow{\cal M} \times \vec{r}}{r^3} \left( 1- \frac{d^2}{2 r^2} (1-5 \frac{z d}{r^2}) + \ldots. \right)
\end{equation}
Note that the $d^0$ term associated to the magnetic charge does not appear. The lowest order non vanishing term has  the conventional structure of the potential of a magnetic moment. The Dirac string singularities of the monopole and the anti-monopole have banished in the expansion. All terms beyond the $d^0$ term are analytic.

Minimal coupling applied to the free Schr\"odinger equation for a spin-less particle of charge $Q$ and mass $m_A$ leads to an interaction  for the magnetic dipole of the form

\begin{equation}
H_{int}= \frac{Q}{2m_A}( \vec{p} \cdot \vec{A} + \vec{A} \cdot \vec{p}) +\frac{ Q^2}{2m_A} \vec{A} \cdot \vec{A}  = H_{par} + H_{dia}.
\end{equation}
Note that $Q g = Z/2$ where $Z e$ is the charge of the particles in the beam. For the velocities involved in of our physical scenario the diamagnetic term, $H_{dia}$, will be small compared to the paramagnetic one. Let us discuss the relation for the lowest order field. The diamagnetic potential for the first term of the potential expansion is

\begin{equation}
~\sim\frac{Z^2 d^2}{2 m_A} \frac{r^2-z^2}{r^6},
 \end{equation}
while the paramagnetic term
\begin{equation}
\sim \frac{Zd v_Q}{r^2} .
\end{equation}
where $v_Q$ is the beam velocity. Being conservative we take for the interesting physical region the following values
 $v_Q\sim1$, $ r\sim \delta $, $r^2-z^2 \sim \delta^2$, where $\delta$ is inter-particle distance in the bunch which is  $\sim 10^7$ fm for protons and $\sim 10^8$ fm for ions and we take for $d$ a maximum value $d \sim 10^3$ fm, which corresponds to the size of a Rydberg monopole-anti-monopole bound state. With these values we get for the ratio of the diamagnetic to the paramagnetic potentials 
 
\begin{equation}
\sim\frac{Z}{A} \frac{d}{2 m_N \delta^2} \sim 10^{-14} - 10^{-12},
\end{equation}
where $m_N$ is the nucleon mass $\sim 1$GeV.

Only for very small velocities and very close to the interaction point is the diamagnetic potential comparable to the paramagnetic one.

In the chosen gauge $\nabla \cdot \vec{A} = 0$, the paramagnetic term can be written as

\begin{equation}
H_{par} = \frac{Q}{m_A} \vec{B} \cdot \vec{L}
\end{equation}
$\vec{B}$ being $\nabla \times \vec{A}$ which is equal to

\begin{equation}
H_{par} = \frac{Q}{m_A} \vec{B} \cdot \vec{L} = \frac{Q}{m_A} \frac{ \overrightarrow{\cal M} \times \vec{r}}{r^3} = - \overrightarrow{\cal M} \cdot \vec{B}_A
\end{equation}
where $\vec{B}_A = Q( \vec{v}_Q \times \vec{r} / r^3)$ is the magnetic field created by the charge in motion. Thus we have shown that within the approximations used the interaction caused by the magnetic field of our magnetic dipole on a charge is the same as the interaction of the magnetic field of the moving charge on the magnetic dipole.

Let us calculate the scattering of charged particles off the magnetic dipole by using the Born approximation which  defines the amplitude for the scattering for a spin-less charged particle by a magnetic dipole as

\begin{equation}
f(\vec{k} \rightarrow \vec{k}') = -4 \pi^2 m_A < \vec{k}' |H_{par}| \vec{k}> = -8 \pi^2 Z e g \int \frac{d^3r}{(2 \pi)^3}    e^{i \vec{k}' \cdot \vec{r}} \left(\frac{\vec{d} \cdot \vec{L}}{r^3} \right)  e^{i  \vec{k} \cdot \vec{r}}.
\end{equation}
In order to have a non vanishing result we take the incoming beam in the $y$ direction, i.e. $\vec{k} =(0, k, 0)$, where k is the incoming momentum. The scattering plane we take as the $xy$-plane, thus $\vec{k^{\prime}} = k (\sin\theta_s, \cos\theta_s,0)$ where $\theta_s$ is the scattering angle.  After some conventional integrations we obtain for the amplitude in the Born approximation from which the cross section becomes

\begin{equation}
 \frac{d\sigma}{d\Omega}(\theta) |_{nr}= Z^2  d^2 \cot^2\theta/2.
 \label{XBorn}
\end{equation}
We see that the cross section is independent of momentum at high energies. 

LHC accelerates particles in bunches which are of macroscopic size $16\mu$m x $16\mu$m x $7.94$cm and contain many particles. Thus the dipole will affect many particles while moving away from the interaction point and separating to distances $d$ up to hundreds of fm. Let us therefore calculate the Born approximation for finite $d$. In principe looking at the expansion this calculation seems prohibitive but  having rewritten the potential as in Eq.(\ref{vectord}) it becomes feasible to do it exactly. Let us apply the Born approximation directly to the full potential in $A_d$, 

\begin{equation}
f(\vec{k} \rightarrow \vec{k}') = - 4 i \pi^2 Z e g \int \frac{d^3 r}{(2 \pi)^3} e^{-i \vec{k}«\cdot \vec{r}} \; \vec{k}\cdot \vec{A}_d \; e^{-i \vec{k}\cdot \vec{r}}.
\end{equation}
We choose as before $\vec{k}= (0,k,0)$ and $\vec{k^{\prime}}= (k \sin \theta_s, k \cos \theta_s,0)$ \footnote{Analogously one could use $\vec{k}=(k,0,0)$ and $\vec{k^{\prime}}= (k \cos \theta_s, k \sin \theta_s,0)$.}.
The calculation requires the following integral

\begin{equation}
\int (\frac{z+d}{\sqrt{x^2+y^2+z^2 +2 z d + d^2}} - \frac{z-d}{\sqrt{x^2+y^2+z^2 -2 z d + d^2}}) \;dz,
\end{equation}
whose result is $4 d$ and is immediate if one recalls the following limit

\begin{equation}
\lim_{z \rightarrow \pm \infty} (\sqrt{x^2+y^2+z^2 +2 z d + d^2} - \sqrt{x^2+y^2+z^2 -2 z d + d^2}) \rightarrow \pm 2 d.
\end{equation}
Having performed this $z$ integral exactly the problem reduces to the simplified calculation performed before and we obtain as result Eq.(\ref{XBorn}),

\begin{equation}
 \frac{d\sigma}{d\Omega}(\theta)|_{nr} = Z^2  d^2 \cot^2\theta/2.
 \label{XBornd}
\end{equation}

Thus we get the same equation for finite $d$ as in the limit $d\rightarrow 0$.

We have performed the calculation in a non-relativistic scheme. Let us now generalise the result by implementing relativistic corrections.

Let us start by studying a beam of spin-less particles. The corresponding Klein-Gordon equation reads
\begin{equation}
((E - A_0)^2 -(\vec{p} -Q \vec{A})^2)\Phi = m^2 \Phi
\end{equation}
Taking $A_0=0$, considering only the paramagnetic interaction term and choosing the gauge where $\vec{\nabla}\cdot\vec{A}=0$
we get
\begin{equation}
(\nabla^2 + k^2) \Phi  = 2 Q (\vec{B} \cdot \vec{L}) \Phi,
\end{equation}
where $k^2 = E^2 - m^2$. This equation has to be compared with the Schr\"odinger equation
\begin{equation}
(\nabla^2 + 2 m E) \Phi  = 2 Q (\vec{B} \cdot \vec{L}) \Phi.
\end{equation}
Thus relativity is implemented just by substituting the non relativistic momentum $k= \sqrt{2 m E}$ by the relativistic one $k =\sqrt{E^2-m^2}$.
Therefore,  the structure of the cross section in the Born approximation  does not change,

\begin{equation}
 \frac{d\sigma}{d\Omega}(\theta) =  \frac{d\sigma}{d\Omega}(\theta)\left|_{nr}\right. .
 \label{dipoleeqKG}
 \end{equation}

Let us assume now that we have a beam of unpolarised  spin $1/2$ particles. Using the conventional notation the Dirac equation for our problem becomes

\begin{equation}
(E+ \vec{\alpha}\cdot (\vec{p}-Q\vec{A}) +m \beta)\Psi =0.
\end{equation}
We next multiply by $E- \vec{\alpha}\cdot (\vec{p}-Q\vec{A}) -m \beta$~\cite{Rose:1948zz,Parzen:1950pa}, and we obtain

\begin{equation}
(\nabla^2 + k^2 ) \Psi  = 2 Q (\vec{B} \cdot \vec{L}) \Psi,
\end{equation}
which leads to the same equation as before for each component using the relativistic momentum. Thus again the structure does not change,

Before closing this section it must be noted that we are performing the calculation in the most favorable situation in which the dipole is perpendicular to the beam.  However, we expect to produce many monopole-anti-monopole pairs which will be created in all possible orientations and therefore the final result will behave as an unpolarised cross section. and will be smaller.

\section{Scattering of charged particles on a pair monopole-anti-monopole}

\begin{figure}[htb]
\begin{center}
\includegraphics[scale= 0.9]{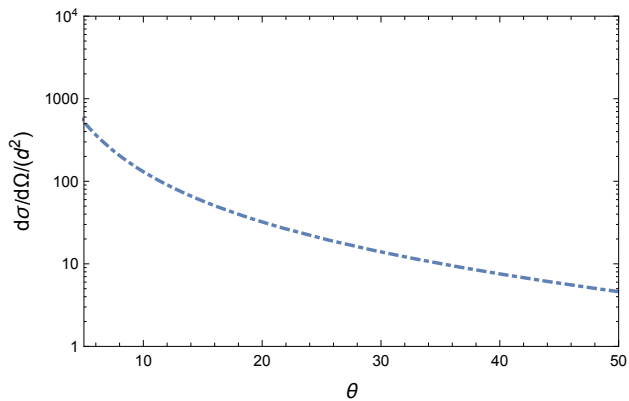} 
\end{center}
\caption{We show the elastic proton-magnetic scattering differential cross section in  Log Plot  in units of $d^2$ for a proton beam as a function of angle. We obtain the typical behaviour of an electromagnetic cross section maximal at small angles decreasing very fast with angle.}
\label{XsecPTheta}
\end{figure}

Let us use the above study for LHC physics. Imagine that monopole-anti-monopole pairs are created in the collisions \cite{Dougall:2007tt,Baines:2018ltl,Epele:2012jn}.  Some of those pairs annihilate into photons and some of them  escape the interaction region. The annihilation cross sections has been studied for some time \cite{Epele:2012jn,Barrie:2016wxf,Fanchiotti:2017nkk}. Those monopoles which escape might be detected directly or bind to matter and methods for detection have been devised \cite{Acharya:2014nyr,Vento:2018sog,Milton:2006cp,Giacomelli:2011re}. We are here interested in discussing what happens while the pairs are escaping the interaction region because this effect might help disentangle the monopole from other exotic particles. A pair of opposite magnetic charges will create a magnetic dipole field as we have shown in the previous section from which the particles of the beam will scatter. We study next what happens with proton and  ion beams at LHC with the maximum planned beam energy $7$ TeV and maximum luminosity.

We study proton beams first. Using Eq.(\ref{XBornd})  we plot in  Fig. \ref{XsecPTheta} the shape of the cross section as a function angle in units of $d^2$. It is a typical electromagnetic cross section large at small angles decreasing rapidly as the angle increases. Thus the wishful signature should occur in the forward direction.

In order to get some realistic estimates for detection we have to fix several scales. The first scale to fix is $d$. The minimum possible value for $d$ is twice the classical radius  of the monopole ($\sim 2 g^2/m$) which for a monopole of mass $500$ GeV is $\sim 0.03$ fm.  For the maximum value of $d$ we choose the separation between monopole-anti-monopole in a magnetic Bohr atom $ (\sim 2 n^2/m g^2)$  for large $n \sim 100$ this leads to  $d \sim 240$ fm. 

The next parameter we need to determine is the duration of the collision. This parameter together with the luminosity of LHC, $2.\; 10^{34} cm^{-2} s^{-1}$, will determine the number of protons scattered by each pair. To determine that number we need to know the maximum separation from the interaction point  at which the dipole is still active and its velocity of separation from the impact point. We will use for the effective separation distance  the width of the bunch $\sim 16 \mu$m, for which we get  $  \beta t \sim 0.8 \,10^{-13}\, s$. In our plots we take for the velocity $\beta$ the value $0.01$, production almost on shell, noting that $\beta$ enters the equation as $\sim 1/\beta$, thus a rescaling of our results is trivial. 

Finally we need to know the number of pairs produced in the collisions. We used the production cross section for spin $0$  monopoles \cite{Dirac:1948um} calculated using the techniques of refs.~\cite{Dougall:2007tt,Baines:2018ltl,Epele:2012jn}.

 Let us discuss first monopoles of 500 GeV mass with Dirac coupling $g$ given by the quantization condition Eq.(1). The cross section for the pairs produced  is $\sim 1000$ pb~\footnote{We are working with Dirac monopoles of charge $g$ thus the cross section is greater than that shown in refs.\cite{Dougall:2007tt,Baines:2018ltl,Epele:2012jn} where they use $\beta$-coupling.}.
 
 \begin{figure}[htb]
\begin{center}
\includegraphics[scale= 0.9]{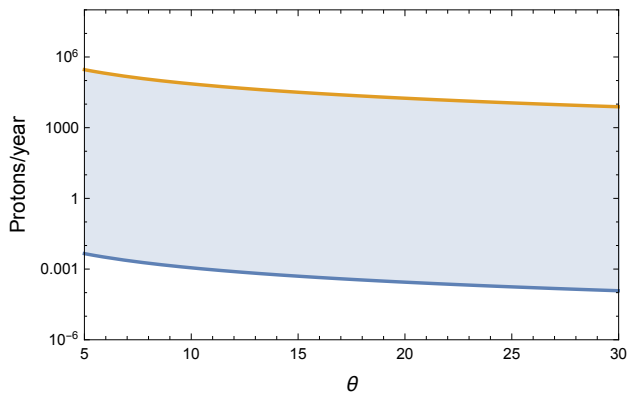} 
\end{center}
\caption{We show the average number of protons scattered in one year from a $7 $ TeV proton beam at an LHC luminosity of  $2\; 10^{34} cm^{-2} s^{-1}$ by a monopole-anti-monopole  separating at  velocity $\beta=0.01$ as a function of scattering angle for the geometry that maximises the cross section. The upper curve corresponds for a dipole of $d\sim 240$ fm and the lower for a dipole of $d \sim 0.03$ fm values which have been justified in the text.}
\label{NumP}
\end{figure}

 With these scales fixed we calculate  the number of protons scattered in one year assuming that the  pair separates with a velocity $\beta=0.01$ from the interaction point. In the process the monopole will separate from the anti-monopole and $d$ will increase from a small value initially to a relatively large value once they leave the proton bunch. The  result of the calculation is shown in Fig. \ref{NumP}. The upper curve corresponds for a dipole of $d\sim 240$ fm and the lower for a dipole of $d \sim 0.03$ fm.The result corresponds to a typical electromagnetic interaction where the forward direction is favoured, but where the non-forward scatterings are an important characteristic. The validity of the Born approximation for the large values of $d$ might be questionable, they have to be taken as an indication of order of magnitude. From an experimental point of view it is the non-forward directions which characterises the creation of a monopole-anti-monopole pair. We see that detection in the near-forward direction is possible for large values of $d$ and the scenario is specially suited for the big detectors ATLAS, CMS and LHCb. These observations  are complementary to direct detection and the annihilation of monopole-anti-monopole pairs into photons. Direct detection might not differentiate monopoles from other exotics and annihilation produces broad  bumps  which are not very characteristic~\cite{Epele:2012jn,Barrie:2016wxf,Fanchiotti:2017nkk}. However, together with the observation of non-forward protons of beam energy these signatures become a clear identification of monopole production.

\begin{figure}[htb]
\begin{center}
\includegraphics[scale= 0.9]{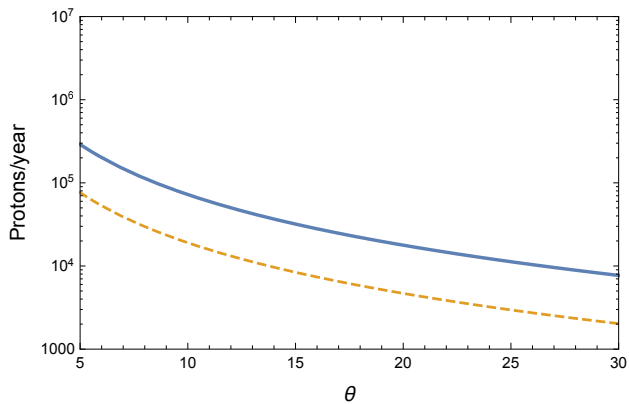} 
\end{center}
\caption{We compare the results obtained previously for the number of protons scattered with Dirac coupling (solid) with those of $\beta$-coupling (dashed) for  $500$ GeV monopole mass. We have taken for this  comparison the largest value of $d=240$ fm.}
\label{NumPbeta}
\end{figure}
 
\begin{figure}[htb]
\begin{center}
\includegraphics[scale= 0.9]{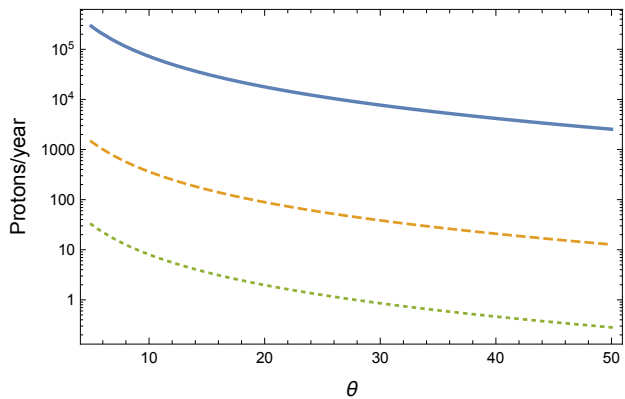} 
\end{center}
\caption{We show the average number of protons scattered in one year  from proton beams at an LHC luminosity of  $2\; 10^{34} cm^{-2} s^{-1}$ by a monopole-anti-monopole  separating at  velocity $\beta=0.01$ as a function of scattering angle for the geometry that maximises the cross section and a maximum $d\sim 240$fm. The curves have been calculated  for three monopole masses: $500$ (solid), $1000$ (dashed), $1500$ (dotted) GeV.}
\label{NumPmass}
\end{figure}

The $\beta$-coupling schemes used in many calculation  \cite{Epele:2012jn,Dougall:2007tt} leads to production cross sections which are smaller and therefore to a smaller number of protons scattered  as shown in Fig. \ref{NumPbeta}. The monopole-anti-monopole production cross section decreases rapidly with the monopole mass \cite{Epele:2012jn,Baines:2018ltl,Dougall:2007tt} and so will the number of scattered protons. We show these results in Fig. \ref{NumPmass} for Dirac coupling. Thus for larger monopole masses the dipole effect becomes more and more difficult to detect. 

\begin{figure}[htb]
\begin{center}
\includegraphics[scale= 0.9]{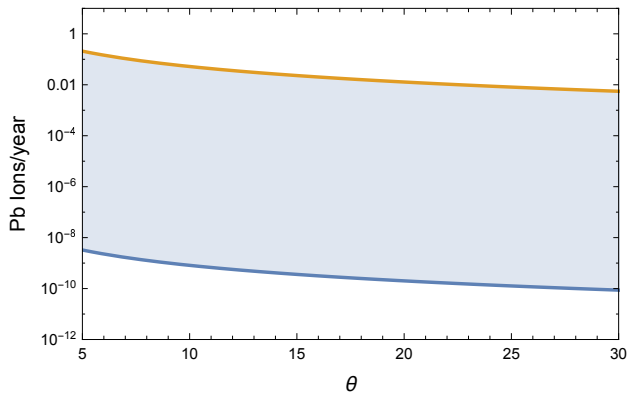} 
\end{center}
\caption{The figure shows the number of ions scattered by a monopole-anti-monopole pair for a $2.76 $ TeV/nucleon  $^{208}Pb^{82+}$ beam, with $\beta=0.01$ and luminosity $10^{27} cm^{-2}s^{-1}$. The upper curve is for  $d\sim 240$fm (solid) and the lower curve for $d=0.03$fm (dashed)}
\label{Num208Pb}
\end{figure}

Let us discuss next heavy ion scenarios by studying $^{208}Pb^{82+}$ beams. In order to get estimates we fix the scales again. The effective duration of the interaction is calculated as before, namely as the time that takes the pair to get out of the bunch. Since the bunches have the same size as for the proton we use the same time scale. We take the same escape velocity of the ions  $\beta=0.01$. Since the collision takes place  in an extreme relativistic scenario we will approximate the lead nucleus by a flat pancake, assume central collisions and thus the production cross section for monopole-anti-monopole pairs can be approximated by $Z^2 \sigma(pp)$, noting that photon fusion is the dominant production mechanism and that the neutrons do not contribute to production. Unluckily the luminosity for ions at LHC is much smaller, $10^{27} cm^{-2}s^{-1}$. This factor proves to be dramatic in not allowing detection. We show in Fig. \ref{Num208Pb} (left) the average number of particles scattered per year for a $^{208}Pb^{82+}$ beam. It is clear that with the present LHC luminosity for lead the possibility of measuring the dipole effect with lead ions is out of question.  In order to see a signature for the MoEDAL detector the luminosity has to be increased minimally by $10^4$. In Fig.\ref{Num208Pb} (right) we show the results for this luminosity. Detection is difficult but possible in the slightly off-forward direction. 

We can summarise the results of our investigation by concluding that the dipole effect of an monopole-anti-monopole pair may be detectable at LHC if monopole masses do not exceed $1000$ GeV with available proton beams and reachable luminosities. With ion beams and present luminosities, detection is not feasible. The signal for the existence of the pair is clear, one should look for protons at beam energies in the off-forward beam directions.

\begin{figure}[htb]
\begin{center}
\includegraphics[scale= 0.9]{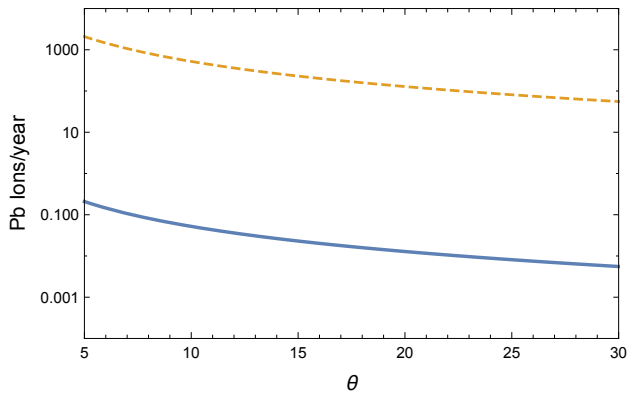} 
\end{center}
\caption{The figure shows the number of ions scattered by a monopole-anti-monopole pair for a $2.76 $ TeV/nucleon  $^{208}Pb^{82+}$ beam, with $d=240$fm , $\beta=0.01$ and luminosity $10^{27} cm^{-2}s^{-1}$ (solid) and  with  $\beta=0.01$ and luminosity $10^{31} cm^{-2}s^{-1}$ (dashed). }
\label{Num208Pb12}
\end{figure}

\section{Scattering off monopolium}
\label{monopolium}

Monopolium is a bound-state of monopole-anti-monopole. It cannot have a permanent dipole moment. However, in the vicinity of a magnetic field it can get an induced magnetic dipole moment through its response to the  external magnetic field. In quantum mechanics the magnetic polarisability $\alpha$ is connected to the change of the energy levels of the system caused by the external field. The general framework to
evaluate these changes is the (stationary) perturbation theory applied to the total Hamiltonian which, in the case of a monopolium 
immersed in a static (and uniform) magnetic field $\vec{B}$,
 can be written 
 \begin{eqnarray}
H({\vec B}) & = & H_0 + H'({\vec B}) = H_0 - {\overrightarrow {\cal M} }\cdot {\vec {\cal B}} \nonumber \\
& = &\frac{{\vec p}\,^2} { 2 \mu} +V_{M \bar M}(r)  - {\cal M}\,{\cal B},
\end{eqnarray}
where $\mu= m/2$ is the reduced mass of the monopole, $V_{M \overline M}$ the potential energy 
associated to the monopole - anti-monopole interaction within a non-relativistic
framework and $\overrightarrow{\cal M}$  the magnetic dipole induced in the system. 
The (negative) lower order correction to the ground state energy of the system
is quadratic in the perturbative field and defines the magnetic (paramagnetic) susceptibility $\alpha_M$

\begin{equation}
\lim_{B \to 0}\,\, \langle{\vec{B}}|\,H_0 + H'({\vec{B}})\,|{\vec{B}} \rangle = E_0 - \frac {1}{ 2}\, \alpha_M \, {\vec{B}}^2  \,\, ,
\label{eq:energy_variation}
\end{equation}
where $|{\vec{B}}\rangle$ is the ground state of
$H({\vec{B}})$, $E_0$ the ground state energy value 
of the unperturbed Hamiltonian $H_0$ and $B=|\vec{B}|$.
The magnetic dipole operator ${\overrightarrow{\cal M}}$ is inferred by the duality from the analogous electric dipole operator:

\begin{equation}
\overrightarrow{D} = e {\vec{r}} \to \overrightarrow{\cal M} = g  \vec{r}
\label{eq:Moperator}
\end{equation}
where ${\vec{r}}$ is the relative position of the monopole and anti-monopole.

The susceptibility $\alpha_M$ can be equivalently defined from
the induced magnetic moment as in the classical case, namely

\begin{equation}
\alpha_M = \lim_{B \to 0}\,\,\frac{\langle {\vec{B}}|{\cal{M}}
|{\vec{B}} \rangle}{ B} \,\, .
\label{eq:aM_definition}
\end{equation}

Both Eqs.~(\ref{eq:energy_variation}) and (\ref{eq:aM_definition}) lead to the well know perturbative
expression (recall that $\langle 0| {{\cal M}}|0\rangle = 0$
because of parity invariance)

\begin{equation}
\alpha_M = 2 \sum_{n \ne 0}^\infty \frac  {| \langle n|{{\cal M}}|0 \rangle|^2}
{E_n - E_0} = 2 m_{-1}({\cal M}),
\label{eq:aM_pert_theory}
\end{equation}
which relates the polarisability $\alpha_M$ to the inverse energy-weighted sum rule $m_{-1}$. Because of the rigorous bounds among sum rules one can estimate
$m_{-1}$ through a lower bound

\begin{equation}
m_{-1}({\cal M}) \ge \frac{m_0^2({\cal M} )}{m_1({\cal M} )}
\end{equation}
with
\begin{eqnarray}
m_0({\cal M})&= &\sum^\infty_{n = 0} |\langle n|{\cal M}|0\rangle |^2 ,\\
m_1({\cal M})&= &\sum^\infty_{n = 0} (E_n-E_0) |\langle n|{\cal M}|0\rangle |^2.
\end{eqnarray}

Thus to get an estimate for the polarisability one has to calculate the first few sum rules for the magnetic dipole
operator (\ref{eq:Moperator}):
\begin{equation}
{\cal{O}} \equiv{{\cal M}} = g\,z,
\end{equation}
where $z$ is the relative distance of monopole and anti-monopole 
in the direction of the external field.
Since monopolium is a two-body system  the calculation of the sum rules can
be performed rather easily not only for the odd moments which depend on
commutators, but for the even moments also, although they require the
evaluation of anticommutators \cite{SR}.


\begin{itemize}

\item[i)] $m_0$ gives the total integrated response
function
\begin{equation}
m_0({{\cal M}}) = \frac{1}{2}\langle 0|\left\{ {{\cal M}}, {{\cal M}} \right\} |0
\rangle  = g^2 \, \frac{1}{ 3}\,\langle 0|{\vec r}^{\;2} |0 \rangle,
\label{eq:m0}
\end{equation}
and is related to the rms radius of the monopolium;


\item[ii)] $m_1$  leads to
\begin{eqnarray}
&& m_1({{\cal M}})  =  \frac{1}{ 2} \langle 0|\left[{{\cal M}}, \left[H_0, {{\cal M}}
\right] \right]|0\rangle = \nonumber \\
& = & g^2 \, \frac {\hbar^2}{ 2 \mu}\,;
\end{eqnarray}
\end {itemize}

The simplicity of the previous commutator is basically
due to the fact that the commonly used monopole-anti-monopole  potentials, $V_{M \overline M}$,  commute with the magnetic dipole operator (\ref{eq:Moperator}).

Lower and upper bounds to the magnetic susceptibility can be found. The so called Feynman bound is given by \cite{traleo94,tra95},

\begin{equation}
\alpha_M \ge \frac{2 m_0^2}{m_1}.
\label{Feynman}
\end{equation}

We make here the assumption that the previous lower bound can reasonably approximate the magnetic susceptibility for monopolium as established in other contexts \cite{SR,traleo94,tra95}, thus

\begin{eqnarray}
\alpha_M &\approx& 2\,\frac{m_0^2({{\cal M}})}{ m_1({{\cal M}})} = \frac{4}{ 9}\,\frac{\mu c^2}{ (\hbar c)^2}\, g^2\, \left[\langle r^2 \rangle \right]^2,
\label{eq:aMlower1}
\end{eqnarray}
where $\langle r^2 \rangle$ is
the mean square radius of the monopolium system. 

\begin{figure}[htb]
\begin{center}
\includegraphics[scale= 0.9]{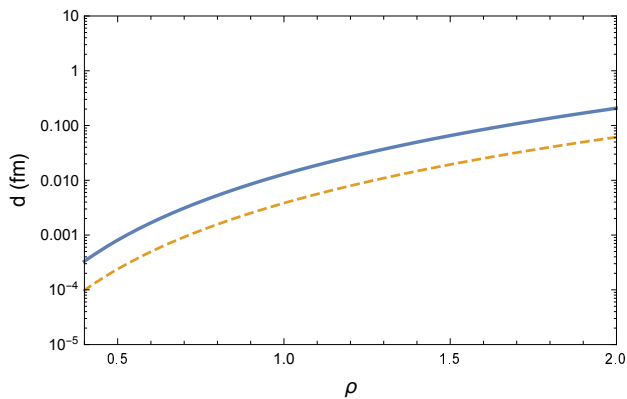} 
\end{center}
\caption{We plot the stiffness distance of monopolium as a function of the size parameter for two values of the monopole mass.}
\label{d}
\end{figure}

Let us describe the physical scenario for detection of monopolium. We assume that monopolium is produced at LHC at 14 TeV fundamentally by photon fusion in the reaction

\begin{equation}
p + p \rightarrow p + p + M.
\end{equation}
Let us assume that monopolium is produced near threshold with a mass below $2000$ GeV. The time scale of the process is dominated by the lifetime of monopolium $t \sim \frac{1}{\Gamma} \sim \frac{1}{10}$ GeV$^{-1}$. The protons travel close to the speed of light and therefore the distance scales are $\sim 0.02$ fm. The magnetic field created by the moving protons  deforms monopolium  and gives it a magnetic moment,

\begin{equation}
\overrightarrow{\cal{M}} = \alpha_M \vec{B} = 2 g \vec{d},
\label{effec:d}
\end{equation}
where we  equate the induced magnetic moment to that of an effective dipole as described in previous sections. $d$ is a measure of the stiffness of monopolium. In this way we will apply the formalism developed in the previous sections to this effective magnetic moment. Our goal is to estimate $\alpha_M$  and $B$ to get $d$ and then we apply the scattering formalism of previous sections.

To calculate $\alpha_M$ we need to have a model for monopolium, i.e. an interaction potential. There are several models in the literature \cite{SchiffGoebel,Barrie:2016wxf} but for the purpose of the present investigation the approximation to the potential of Schiff and Goebel \cite{SchiffGoebel} 

\begin{equation}
V(r) = -g^2 \frac{1-exp(-2 r/r_0)}{r},
\end{equation}
used in ref.\cite{Epele:2007ic} will be sufficient. The approximation consists in substituting the true wave functions by Coulomb wave functions of high $n$. For each $r_0$ a different value of large $n$ will be best suited. We use the equation

\begin{equation}
\rho= 48 \alpha^2  n^2 
\label{size}
\end{equation}
to parametrise all expectation values in terms of $\rho$, where $\rho=r_M/r_{classical}$, $r_M$ being the expectation value of $r$ in the $(n,0)$ Coulomb state, and $\alpha$ the electromagnetic fine structure constant $\sim \frac{1}{137}$. We allow $\rho$ to be continuous parameter representing in such a way potentials of different cutoff ranges. In terms of rho the binding energy becomes

\begin{equation}
M=  m(2-\frac{3}{4 \rho}),
\label{binding}
\end{equation}
a function in terms of $\rho$ which covers the interval $[0, 2m]$.
In this approximation all the moments can be determined analytically
\begin{eqnarray}
m_0 & = & \frac{1}{864} \frac{\rho (5 \rho + 48 \alpha^2)}{m^2}, \nonumber \\
m_1 & = & \frac{1} {4 \alpha m}, \nonumber 
\end{eqnarray}
The magnetic susceptibility obtained from the Feynman estimate Eq.(\ref{Feynman}) becomes
\begin{equation}
\alpha_M = \frac{1}{93312} \frac{\rho^2 (5 \rho +48 \alpha^2)^2}{m^3 \alpha^5}.
\end{equation}

\begin{figure}[htb]
\begin{center}
\includegraphics[scale= 0.9]{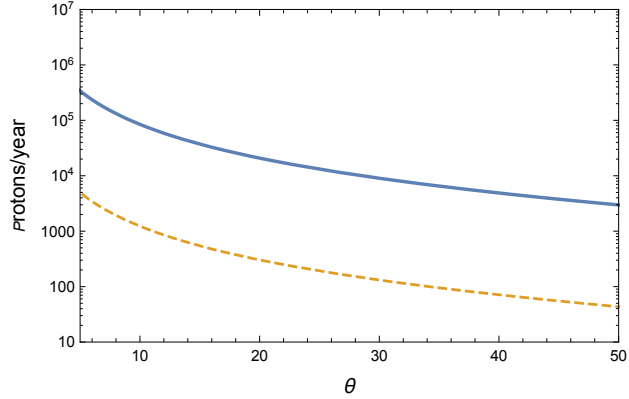} 
\end{center}
\caption{We plot the number of protons scattered by monopolium as a function of scattering angle in the strong binding limit for a monopole mass of $m= 1000$ GeV and range parameter $\rho =0.5$. The upper curve corresponds to Dirac coupling and the lower curve to $\beta$-coupling.}
\label{NMbeta}
\end{figure}

Let us now estimate the magnetic field acting on monopolium. We are assuming that monopolium is moving slowly as compared to the protons ($\beta_M \sim 0.01)$ , therefore it is static in the time scale of the problem. The proton which creates the magnetic field is moving very fast, $\beta_p \sim 1$. The time scale of the problem is determined by the lifetime of monopolium which leads to an effective radius of $R \sim 0.02$ fm, thus

\begin{equation}
B \sim \frac{2 e \beta_p}{R^2} \sim 8.5 \;\mbox{GeV}^2.
\end{equation}
The effective distance, Eq. (\ref{effec:d}) becomes
\begin{equation}
d= 0.15 \left( \frac{\alpha_M}{\mbox{GeV}^3}\right)\;  \mbox{fm}
\end{equation}
In order to perform the calculation we require the monopolium production cross section. To do so we have used the formalism and computational programs of \cite{Epele:2007ic} with updated pdfs. 

In Fig. \ref{d} we plot $d$ as a function of size parameter for $m = 1000$ GeV and $m= 1500$ GeV. We note that the stiffness parameter ranges from 0.001 fm for strong bound monopolium to 0.1 fm for weak bound monopolium. 

\begin{figure}[htb]
\begin{center}
\includegraphics[scale= 0.8]{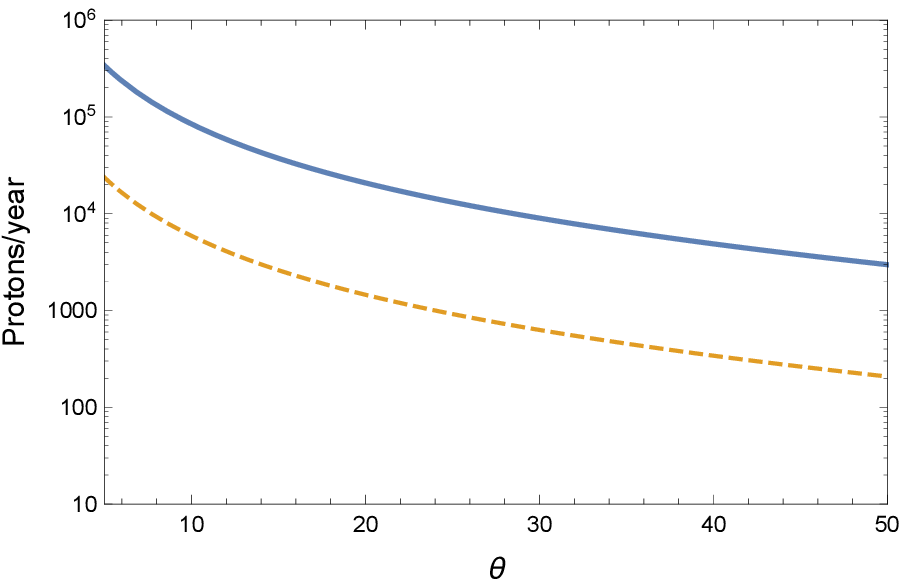} \hspace{1.0cm} \includegraphics[scale= 0.8]{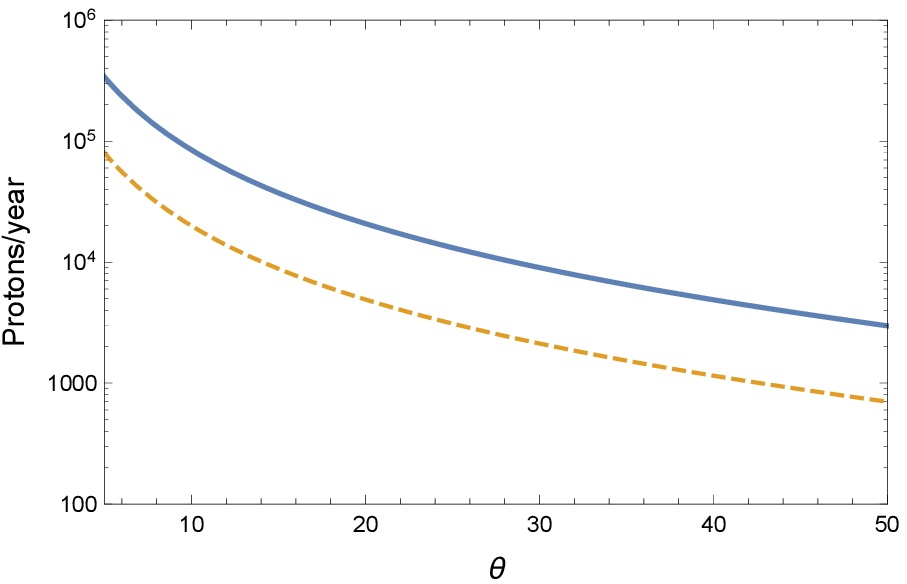} \end{center}
\caption{We show the average number of protons scattered in one year from a proton beam at an LHC luminosity of  $2\; 10^{34} cm^{-2} s^{-1}$ by monopolium moving at  velocity $\beta=0.01$ as a function of scattering angle for the geometry that maximises the cross section. Left: for $k=7$ TeV, $m=1000$ GeV and $\rho=0.5$ (solid) and $\rho=1$ (dashed) ; Right: for $\rho =0.5$, $k= 7$ TeV, $m= 1000$ GeV (solid) and $m=1500$ GeV (dashed).}
\label{NM}
\end{figure}

We limit here the calculation to proton beams since we have seen in the previous sections that the luminosity for ions is too low to produce detectable results. In Figs. \ref{NM} we show the dependence of the number of protons scattered per year as a function of scattering angle for different values of the size parameter and  monopole masses. The figure on the left shows that the strong binding scenario might allow detection, for monopolia with a mass below $1000$ GeV, while observations in the weak binding scenario are difficult. This has to do with the monopolium production cross section which decreases very fast with the $\rho$ parameter compensating for the smaller stiffness. The figure in the right shows that as we increase the monopole mass for a fixed $\rho$  the cross sections becomes smaller and observability is reduced.   The production cross section diminishes greatly as the monopole mass increases.

We have not studied here the energy dependence of the cross section since in the Born approximation the cross section comes out energy independence and all the energy dependence will come from the production cross section \cite{Epele:2012jn,Epele:2007ic}. We have presented all results for $7$ TeV proton beams and LHC luminosities.

To summarise, we stress that the analysis for monopolium depends strongly on the details of the dynamics. Different monopole-anti-monopole potentials might lead to different results. In particular, the phenomenon depends very strongly on the binding energy and the decay width. Large binding energies and small widths will increase observability. Given the neutral nature of monopolium the detection of non-forward protons is ideal for its characterisation. However, in weakly bound monopolia with short lifetimes given the planned luminosities at LHC the phenomenon would not be observable.

\section{Concluding remarks}

In previous work we studied ways to detect Dirac monopoles bound in matter by means of proton and ion beams \cite{Vento:2018sog}. We also  studied the possibility of finding monopoles not free but in bound pairs of monopole-anti-monopole, the so called monopolium \cite{Epele:2012jn,Epele:2007ic}. Monopolium has lower mass than a pair of monopole-anti-monopole and also annihilates into photons \cite{Barrie:2016wxf,Fanchiotti:2017nkk} but because it is neutral it is difficult to detect directly. In this paper we pursue some investigations to detect monopoles in LHC besides direct detection and their decay properties. We study the distortion produced in the beam by their permanent or induced magnetic dipole moment to characterise detectability. We have modelled the interaction  by a fixed magnetic dipole made by two magnetic charges $g$ and $-g$ separated by a distance $d$. We have studied how this effective magnetic dipole interacts with a beam of charged particles. The main result is that the beam particles will be deflected and therefore particles with beam energy will appear in off-forward directions. We have shown that monopole-anti-monopole pairs lead to an sizeable effect with the proton beams at LHC and thus the effect is suitable for detection in ATLAS, CMS and LHCb. However, present heavy ion luminosities do not allow detection which makes the scenario not useful for MoEDAL. In the case of monopolium the strong coupling limit also leads to off-forward protons, a scenario which could characterise the production of this neutral particle. However our study  shows that observability of the phenomenon depends very strongly on the lifetime of monopolium and its binding energy.

To conclude monopoles  can be detected directly  or by the decay of monopole-anti-monopole pairs into photons. Monopolium can  be detected by its decay into photons.  We have shown that detecting beam particles at  beam energy in non-forward directions  becomes an additional tool for monopole and monopolium detection.

\section*{Acknowledgement}

M.T. thanks the Department of Theoretical Physics of 
the University of Valencia for a Visiting Professor grant and for the warm and friendly hospitality. VV acknowledges fruitful discussions with Jos\'e Bernab\'eu.
This work was supported in part  
by the MICINN and UE Feder under contract FPA2016-77177-C2-1-P and SEV-2014-0398.

\end{document}